\documentclass[11pt]{article}

\usepackage[utf8]{inputenc}
\usepackage[T1]{fontenc}
\usepackage{lmodern}
\usepackage{amsmath,amssymb,amsfonts}
\usepackage{graphicx}
\usepackage{cite}
\usepackage{hyperref}
\usepackage{geometry}
\usepackage{algorithm}
\usepackage{algpseudocode}
\usepackage{xurl}

\title{A Correlation Aware Quantum Feature Map for Variational Quantum Classification}

\author{
Murat Kurt\\
\small Department of Software Engineering, Samsun University, Samsun, Türkiye\\
\small \texttt{kmurat.physics@gmail.com}
}

\begin{document}

\maketitle

\begin{abstract}
Quantum machine learning has emerged as a promising research area for learning complex data patterns. However, most existing quantum feature maps employ fixed encoding strategies that do not explicitly consider the relationships among features within a dataset. In this study, we propose a Correlation Aware Quantum Feature Map (CAQFM) which integrates feature dependencies into the quantum encoding process. The proposed approach utilizes Pearson, Spearman, Kendall Tau, Mutual Information, and Distance Correlation measures to identify relationships among features. Dependencies exceeding a predefined threshold are incorporated into the quantum circuit through controlled quantum gates, enabling the construction of richer quantum representations that better reflect the underlying structure of the data.

The proposed method is evaluated using a Variational Quantum Classifier (VQC) on three benchmark datasets, namely breast cancer diagnosis, credit default prediction, and student placement classification. Simulation results demonstrate that correlation based quantum encoding can improve classification performance compared to conventional encoding strategies. In particular, the Spearman and Kendall Tau based CAQFM variants achieved the highest predictive performance and consistently outperformed standard quantum feature maps. The findings indicate that incorporating dependency information from classical data into quantum feature maps facilitates the generation of more discriminative quantum representations and enhances the effectiveness of variational quantum classifiers.
\end{abstract}

\noindent\textbf{Keywords:} Quantum Machine Learning, Quantum Feature Map, Variational Quantum Classifier, Correlation Aware Encoding, Quantum Neural Networks

\section{Introduction}

Quantum computing continues to evolve as a computational paradigm with the potential to provide significant computational and security advantages over classical computing for certain classes of problems by exploiting quantum mechanical phenomena such as superposition, interference, entanglement, and the no-cloning principle\cite{benioff1980,feynman1982,deutsch1985,bennett1995}. With the development of Noisy Intermediate-Scale Quantum (NISQ) hardware architectures, research on applying quantum computing to machine learning problems has accelerated considerably\cite{preskill2018,AbuGhanem2023}. A substantial portion of these efforts has focused on the field of Quantum Machine Learning (QML), which aims to combine the representational power and parallel processing capabilities of quantum systems with modern learning algorithms\cite{lohia2024,klusch2024,acampora2025}. One of the most critical components determining the success of QML models is how classical data are encoded into quantum states. Since most real-world datasets are inherently classical, they must be transformed into suitable quantum representations before being processed by quantum algorithms. To this end, various encoding strategies, including angle encoding, amplitude encoding, and basis encoding, have been proposed \cite{Sharma2024,rath2024,khan,weigold}. More recently, feature map-based approaches such as Z Feature Map, ZZ Feature Map, and Pauli Feature Map have become widely adopted in variational quantum classifiers and quantum neural networks \cite{Singh2025,suzuki2024,ozpolat2024,bartkiewicz2020,toufah2025,Alami2024}. Although existing quantum feature maps have demonstrated promising performance across various classification tasks, they often fail to account for the underlying statistical and relational structure of the data. In particular, dependencies among features are generally ignored, and interaction patterns within quantum circuits are typically defined independently of the dataset itself \cite{Hirai2024,ArthurDate2022,Tyrovolas2024,Idzikowski2026,MurrayParsons2024,HowCheah2024,BuresFrantis2024}. Consequently, the resulting quantum representations may capture only a limited portion of the structural information embedded in the data.

In both classical machine learning and QML, dependencies among features play a fundamental role in understanding and modeling complex data structures. Accordingly, statistical measures such as Pearson Correlation, Spearman Correlation, Kendall Tau Coefficient, Mutual Information, and Distance Correlation are widely employed to identify linear and nonlinear relationships between variables \cite{Spearman1904,deWinter2016,Stepanov2015O,TisocBeltran2022,Edelmann2021}. These dependencies often reveal hidden structures within datasets and can directly influence learning performance. Nevertheless, studies that systematically incorporate feature dependencies into quantum data encoding processes remain relatively scarce.

To address this limitation, this study proposes a Correlation-Aware Quantum Feature Map (CAQFM) that directly incorporates feature dependencies into the quantum data encoding process. In the proposed approach, relationships between feature pairs are quantitatively characterized using Pearson Correlation, Spearman Correlation, Kendall Tau Coefficient, Mutual Information, and Distance Correlation. Dependency values exceeding a problem-specific threshold are subsequently used to determine the rotation angles of controlled quantum gates. As a result, the architecture of the quantum encoding circuit is adaptively shaped according to the statistical characteristics of the dataset.

The proposed method is evaluated on three binary classification problems: breast cancer diagnosis, credit default prediction, and student placement prediction. Experimental results demonstrate that incorporating relational information into the quantum encoding process can improve classification performance compared with conventional feature maps employing fixed interaction structures such as linear, circular, and fully entangling connectivity patterns. Furthermore, the proposed approach establishes an interpretable connection between classical dependency analysis and quantum data representation, providing a new perspective for the development of data-driven quantum feature maps.

\section{Related Work}

\subsection{Variational Quantum Algorithms}

With the development of NISQ devices, research on quantum machine learning has increased significantly. Due to the limited number of qubits available in current quantum hardware and their susceptibility to noise, the practical implementation of large-scale fault-tolerant quantum algorithms remains challenging. Consequently, researchers have increasingly focused on hybrid approaches that can exploit the capabilities of existing quantum devices. One of the most prominent examples of such approaches is Variational Quantum Algorithms (VQAs). In VQAs, a parameterized quantum circuit is executed on a quantum processor, while the circuit parameters are iteratively updated by a classical optimizer. In this manner, quantum computation and classical optimization are combined to perform learning and optimization tasks \cite{Cerezo2021,Qi2024,Endo2020,Wu2021}.

Among VQA-based models, Variational Quantum Classifiers (VQCs) and Quantum Neural Networks (QNNs) are the most widely used approaches in the field of Quantum Machine Learning (QML). These models generally consist of four main components: a data encoding layer, a variational circuit (ansatz), a measurement layer, and an optimization procedure. The data encoding layer transforms classical data into quantum states, while the variational circuit contains trainable parameters. The measurement layer extracts classical information from the quantum system, and the resulting outputs are used by the optimization process to update the circuit parameters iteratively \cite{Qin2023,danyal,Zhao2021,Schuld2014,Beer2020}.

The performance of VQA-based models largely depends on the chosen ansatz architecture and optimization strategy. Hardware-efficient ansatz designs are among the most commonly employed architectures in current implementations. These circuits typically combine parameterized single-qubit rotation gates with entangling gates such as CNOT operations. For parameter optimization, algorithms such as COBYLA, SPSA, and gradient-based methods are frequently utilized. Recent studies have demonstrated that VQC and QNN models can achieve promising results across a wide range of applications, including healthcare, finance, image processing, and pattern recognition.

\subsection{Encoding and Feature Maps}

The transformation of classical data into quantum states is a fundamental step in quantum machine learning. Unlike classical computers, which operate directly on classical data vectors, quantum computers process information encoded in quantum states. Therefore, classical data must be mapped into suitable quantum representations before quantum algorithms can be applied. To address this requirement, several quantum data encoding strategies have been proposed. Basis encoding represents information using computational basis states, while amplitude encoding stores information within the amplitudes of quantum states. Angle encoding, on the other hand, maps feature values to the rotation angles of quantum gates such as \(R_X\), \(R_Y\), and \(R_Z\). Due to its simplicity and compatibility with current quantum hardware, angle encoding has become one of the most widely adopted encoding techniques in VQAs. In addition to these approaches, feature map-based encoding methods have emerged as an important research direction in recent years. Feature maps aim to project classical data into higher dimensional Hilbert spaces, thereby generating richer quantum representations. Such representations can enable quantum classifiers to extract meaningful patterns from more complex data structures.

One of the most widely used feature maps is the Z Feature Map, in which classical features are encoded into quantum states through phase transformations. An extension of this approach, known as the ZZ Feature Map, introduces pairwise interactions between qubits, allowing certain relationships among features to be represented within the quantum circuit. To further enhance representational capacity, the Pauli Feature Map employs combinations of different Pauli operators to generate more expressive quantum states.

These feature map-based approaches have demonstrated promising performance in a variety of quantum classification tasks and have become standard data encoding techniques in QML. However, their interaction structures are generally predefined and remain independent of the statistical properties of the underlying dataset. Consequently, the relational information present among features may not be fully reflected in the resulting quantum representations.

\subsection{Limitations of Existing Quantum Feature Maps}

Although existing quantum feature maps have demonstrated promising performance across a variety of classification tasks, they exhibit several important limitations. One of the primary limitations is that the interaction structures employed within quantum circuits are generally designed without considering the relative importance or dependency patterns among features in the dataset. In many studies, linear, circular, or fully connected entanglement topologies are adopted and applied uniformly across different datasets. This approach implicitly assumes that all features contribute equally to the learning process. However, in real-world applications, some features may exhibit strong relationships, whereas others may be only weakly correlated or entirely independent. When fixed interaction structures are used, these differences are ignored, and both strong and weak dependencies are treated identically.

Another important limitation is that most existing feature maps primarily focus on encoding individual feature values. In classical machine learning, relationships among variables often reveal the underlying structure of a dataset and can significantly influence model performance. Nevertheless, such relational information is rarely incorporated into the quantum data encoding process. Furthermore, the use of fixed interaction topologies may introduce unnecessary quantum gates into the circuit. Considering the noise and decoherence challenges associated with current NISQ devices, an excessive number of gates can increase circuit depth and consequently degrade model performance.

These observations indicate an important research gap in the QML literature. While the majority of existing studies focus on developing more sophisticated feature maps or more expressive variational circuits, relatively few approaches directly incorporate statistical relationships within the dataset into the quantum data encoding process. To address this limitation, this work proposes the Correlation-Aware Quantum Feature Map (CAQFM). In the proposed approach, dependencies among features are extracted directly from the dataset and utilized to determine the structure of controlled quantum rotation gates. As a result, the quantum data encoding process can be adaptively tailored to the statistical characteristics of the underlying dataset.

\section{Proposed Method}

In this section, the proposed Correlation-Aware Quantum Feature Map (CAQFM), which incorporates dependencies among features into the quantum data encoding process, is presented in detail. The proposed approach establishes a direct connection between classical data analysis and quantum data encoding, enabling the statistical structure of a dataset to be embedded into quantum representations. Within the proposed QML framework, this procedure is employed as a classical preprocessing stage prior to quantum encoding. By extracting dependency information from the dataset before constructing the quantum circuit, the encoding process can be adaptively tailored to the underlying characteristics of the data.

\subsection{Types of Correlation}

The first stage of CAQFM consists of quantifying dependencies among the features of a dataset. To this end, several dependency measures, including Pearson Correlation, Spearman Correlation, Kendall Tau Coefficient, Mutual Information, and Distance Correlation, are employed.

Pearson correlation measures the strength of the linear relationship between two variables and is defined as

\begin{equation}
r_{xy}
=
\frac{
\sum_{k=1}^{n}
(x_k-\bar{x})
(y_k-\bar{y})
}
{
\sqrt{
\sum_{k=1}^{n}
(x_k-\bar{x})^2
}
\sqrt{
\sum_{k=1}^{n}
(y_k-\bar{y})^2
}
}
\label{eq:pearson}
\end{equation}

where \(x_k\) and \(y_k\) denote the \(k\)-th observations of two features, and \(\bar{x}\) and \(\bar{y}\) represent their corresponding sample means. The coefficient takes values in the interval \([-1,1]\), where values close to \(+1\) indicate a strong positive linear relationship, values close to \(-1\) indicate a strong negative linear relationship, and values near zero indicate little or no linear dependence.

Spearman correlation is a non-parametric measure that quantifies the strength of a monotonic relationship between two variables. Unlike Pearson correlation, it operates on ranked observations rather than the original values, making it suitable for capturing nonlinear monotonic dependencies and more robust against outliers.

The Spearman correlation coefficient is computed as

\begin{equation}
\rho
=
1-\frac{6\sum_{i=1}^{n} d_i^2}{n(n^2-1)}
\label{eq:spearman}
\end{equation}

where \(n\) denotes the number of observations and \(d_i\) represents the difference between the ranks of the corresponding observations. The coefficient ranges from \(-1\) to \(+1\), where positive values indicate increasing monotonic relationships and negative values indicate decreasing monotonic relationships.

Kendall Tau is another non-parametric dependency measure based on the difference between concordant and discordant observation pairs. It is defined as

\begin{equation}
\tau
=
\frac{N_c-N_d}{\frac{n(n-1)}{2}}
\label{eq:kendall}
\end{equation}

where \(N_c\) and \(N_d\) denote the numbers of concordant and discordant pairs, respectively. Similar to Spearman correlation, Kendall Tau takes values in the interval \([-1,1]\).

Mutual Information (MI) is employed to quantify both linear and nonlinear dependencies between variables. Unlike correlation-based measures, MI evaluates the amount of shared information between two variables and is defined as

\begin{equation}
I(X;Y)
=
\sum_{x \in X}
\sum_{y \in Y}
p(x,y)
\ln
\frac{p(x,y)}{p(x)p(y)}
\label{eq:mi}
\end{equation}

where \(p(x,y)\) denotes the joint probability distribution of \(X\) and \(Y\), while \(p(x)\) and \(p(y)\) represent their marginal probability distributions.

Distance Correlation is a more general dependency measure capable of detecting both linear and nonlinear relationships between variables. Unlike Pearson correlation, it can identify any form of statistical dependence and is therefore particularly useful for analyzing complex datasets.

Distance Correlation is defined as

\begin{equation}
dCor(X,Y)
=
\frac{dCov(X,Y)}
{\sqrt{dVar(X)\,dVar(Y)}}
\label{eq:dcor}
\end{equation}

where \(dCov(X,Y)\) denotes the distance covariance between variables \(X\) and \(Y\), and \(dVar(X)\) and \(dVar(Y)\) denote their corresponding distance variances. The resulting coefficient ranges from \(0\) to \(1\), where larger values indicate stronger dependencies.

\subsection{Dependency Matrix Construction}

Using one of the dependency measures described above, pairwise relationships between all features are calculated to construct a dependency matrix representing the relational structure of the dataset:

\[
C=
[c_{ij}]
\]

where \(c_{ij}\) denotes the dependency coefficient between the \(i\)-th and \(j\)-th features. Each element of the matrix quantifies the strength of the relationship between a pair of features. Larger absolute values correspond to stronger dependencies.

Encoding all pairwise relationships into a quantum circuit may introduce an excessive number of controlled gates and unnecessarily increase circuit depth. This issue is particularly critical for NISQ devices, where deeper circuits are more susceptible to noise and decoherence. Therefore, a threshold parameter \(\tau\) is introduced to retain only statistically significant relationships.

The corresponding adjacency matrix is defined as

\begin{equation}
A_{ij}
=
\begin{cases}
1, & |c_{ij}| \ge \tau \\[4pt]
0, & |c_{ij}| < \tau
\end{cases}
\label{eq:adjacency}
\end{equation}

where \(A_{ij}=1\) indicates that the corresponding feature pair will be connected through a controlled quantum rotation gate, while \(A_{ij}=0\) indicates that no quantum interaction will be introduced between the two features.

As a result, the thresholding procedure enables the selection of the most informative dependencies within the dataset, leading to a more data-driven quantum encoding architecture. This strategy not only reduces unnecessary quantum operations but also allows the statistical structure of the dataset to be more effectively embedded into the resulting quantum representation.

\subsection{CAQFM Circuit Architecture}

The proposed CAQFM architecture consists of three sequential stages: a quantum encoding layer, a correlation-aware interaction layer, and a variational learning layer. Figure~X illustrates the overall structure of the proposed framework.

In the first stage, each normalized feature is encoded into a corresponding qubit using angle encoding. Given a feature vector

\[
\mathbf{x}
=
(x_1,x_2,\ldots,x_n),
\]

the initial quantum state is prepared as

\[
|\Psi_{enc}\rangle
=
\bigotimes_{i=1}^{n}
R_Y(x_i)|0\rangle .
\]

This operation embeds the classical feature vector into the Hilbert space while preserving the dimensional structure of the original data.

Following the encoding stage, the correlation-aware interaction layer is applied. Let

\[
C=[c_{ij}]
\]

denote the dependency matrix obtained using one of the selected statistical dependency measures. After thresholding, only feature pairs satisfying

\[
|c_{ij}| \ge \tau
\]

are retained. For each selected feature pair, a controlled rotation gate is introduced into the quantum circuit. In this work, controlled \(R_Y\) gates are employed:

\[
CRY(\theta_{ij}).
\]

The rotation angle is determined directly from the corresponding dependency coefficient,

\[
\theta_{ij}
=
\frac{\pi}{2} |c_{ij}|,
\]

allowing stronger feature dependencies to induce stronger quantum interactions. Consequently, the connectivity pattern of the quantum circuit is not predefined but is adaptively generated from the statistical structure of the dataset.

The resulting state is subsequently processed by a variational learning layer consisting of parameterized single-qubit rotations

\[
R_X(\alpha_i),
\qquad
R_Y(\beta_i),
\qquad
R_Z(\gamma_i),
\]

followed by a linear chain of entangling CNOT gates,

\[
CX(q_i,q_{i+1}).
\]

This hardware-efficient ansatz introduces trainable parameters while maintaining compatibility with the connectivity constraints of current NISQ devices.

Finally, all qubits are measured in the computational basis and the resulting probability distribution is used to perform the classification task.

\subsection{Training Procedure}

The proposed CAQFM-based quantum classifier is trained using a hybrid quantum-classical learning framework. In this approach, quantum state preparation and measurement are performed on a quantum processor, while the trainable parameters of the variational circuit are optimized by a classical optimizer. This hybrid strategy enables efficient learning while remaining compatible with the limitations of current Noisy Intermediate-Scale Quantum (NISQ) devices.

For binary classification tasks, the Binary Cross-Entropy (BCE) loss function is employed:

\begin{equation}
L
=
-\frac{1}{N}
\sum_{i=1}^{N}
\Bigl[
y_i \log(p_i)
+
(1-y_i)\log(1-p_i)
\Bigr]
\label{eq:bce}
\end{equation}

where \(N\) denotes the number of samples, \(y_i\) is the true class label, and \(p_i\) represents the predicted probability produced by the quantum classifier.

The optimization of the variational parameters is performed using the Constrained Optimization BY Linear Approximation (COBYLA) algorithm. COBYLA is a gradient-free optimization method that has been widely adopted in variational quantum algorithms due to its robustness and suitability for noisy quantum environments.

The overall construction and training procedure of the proposed CAQFM framework is summarized in Algorithm~\ref{alg:caqfm}.

\begin{algorithm}[t]
\caption{Correlation-Aware Quantum Feature Map Training Procedure}
\label{alg:cefm}

\begin{algorithmic}[1]
\Require Dataset $D=\{(x_i,y_i)\}_{i=1}^{N}$, dependency measure $M$, threshold $\tau$, maximum iterations $T$
\Ensure Optimized variational parameters $\Theta^{*}$

\State Normalize all features in $D$
\State Compute pairwise dependencies using measure $M$
\State Construct dependency matrix $C$
\State Generate adjacency matrix according to threshold $\tau$
\State Build the correlation-aware quantum feature map
\State Initialize variational parameters $\Theta$

\For{$t=1$ to $T$}
    \State Encode classical data into quantum states
    \State Execute the variational quantum circuit
    \State Measure output probabilities
    \State Compute Binary Cross-Entropy loss
    \State Update $\Theta$ using COBYLA optimizer
\EndFor

\State \Return $\Theta^{*}$

\end{algorithmic}
\end{algorithm}

The training procedure consists of two main stages. First, the CAQFM structure is generated from the statistical dependencies extracted from the dataset. The resulting dependency graph determines the placement and strength of controlled quantum interactions within the feature map. Second, the parameters of the variational quantum classifier are optimized using the Binary Cross-Entropy objective and the COBYLA optimizer. This framework enables feature values and feature dependencies to be jointly incorporated into the quantum encoding process, resulting in a data-driven quantum representation that reflects the statistical characteristics of the underlying dataset.

\section{Experimental Setup}

This section describes the datasets, preprocessing procedures, experimental configuration, and evaluation metrics used to assess the performance of the proposed CAQFM framework.

\subsection{Datasets}

To evaluate the effectiveness of the proposed approach, three binary classification datasets from healthcare, finance, and education domains were employed. These datasets were selected to investigate the robustness of CAQFM across different application areas.

\begin{table}[H]
\centering
\caption{Datasets used in this study}
\label{tab:datasets}
\begin{tabular}{cccc}
\hline
\textbf{Dataset} & \textbf{Samples} & \textbf{Features} & \textbf{Classes} \\
\hline
Breast Cancer & 569 & 4 & 2 \\
Credit Default & 566 & 3 & 2 \\
Student Placement & 2000 & 4 & 2 \\
\hline
\end{tabular}
\end{table}

\subsection{Data Preprocessing and Feature Selection}

Missing-value analysis and data cleaning procedures were first applied to all datasets. Class labels were subsequently converted into numerical form to facilitate binary classification. To eliminate the effects of differing feature scales, all features were standardized prior to model training.
To mitigate potential bias caused by class imbalance, balanced datasets were generated before training. Finally, feature selection was performed to reduce the number of required qubits and retain the most informative variables for each dataset.

\begin{table}[H]
\centering
\caption{Selected features from data sets}
\label{tab:features}
\footnotesize
\begin{tabular}{lcccc}
\hline
\textbf{Dataset} & \textbf{F1} & \textbf{F2} & \textbf{F3} & \textbf{F4} \\
\hline
Breast Cancer & CP-worst & Perim-worst & CP-mean & Rad-worst \\
Credit Default & Age & Loan & Income & -- \\
Student Placement & Exam & Study hrs & Assignments & Prev. score \\
\hline
\end{tabular}

\vspace{2mm}
\begin{flushleft}
\footnotesize
\textit{Note.} CP: concave points; Perim: perimeter; Rad: radius; 
Prev.: previous; hrs: hours.
\end{flushleft}
\end{table}

\subsection{Experimental Configuration}

The proposed CAQFM framework was evaluated within a Variational Quantum Classifier (VQC) architecture. Classical features were encoded into quantum states using \(R_Y\) rotation gates, while feature dependencies identified by Pearson Correlation, Spearman Correlation, Kendall Tau Correlation, Mutual Information, and Distance Correlation were incorporated into the quantum circuit through controlled \(R_Y\) rotation gates.

A single-layer hardware-efficient variational ansatz was employed in all experiments. The dataset was divided into training and test subsets using a fixed test ratio of 0.25. Model parameters were optimized using the COBYLA optimizer with an initial trust-region radius parameter of \texttt{rhobeg}=0.5. Binary Cross-Entropy was adopted as the objective function during training.

Since the statistical characteristics of each dataset differ, the dependency threshold parameter \(\tau\) was selected separately for each problem. A threshold value of \(0.95\) was used for the Breast Cancer dataset, whereas thresholds of \(0.20\) and \(0.10\) were employed for the Credit Default and Student Placement datasets, respectively.

The experimental settings used throughout this study are summarized in Table~\ref{tab:expconfig}.

\begin{table}[H]
\centering
\caption{Experimental configuration}
\label{tab:expconfig}
\begin{tabular}{lc}
\hline
\textbf{Parameter} & \textbf{Value} \\
\hline
Train/Test Split & 75\% / 25\% \\
Ansatz Layers & 1 \\
Optimizer & COBYLA \\
COBYLA rhobeg & 0.5 \\
Loss Function & Binary Cross-Entropy \\
Breast Cancer Threshold (\(\tau\)) & 0.95 \\
Credit Default Threshold (\(\tau\)) & 0.20 \\
Student Placement Threshold (\(\tau\)) & 0.10 \\
\hline
\end{tabular}
\end{table}

To assess classification performance, Accuracy, Precision, Recall, F1-score, and ROC-AUC metrics were computed for all experiments.

\section{Results and Discussion}

In this section, the performance of the proposed CAQFM framework is evaluated on three different datasets. The obtained results are compared with conventional quantum feature maps. For all simulations, Accuracy, Precision, Recall, F1-score, and ROC-AUC metrics are reported.

\subsection{Breast Cancer Dataset Results}

The Breast Cancer Wisconsin dataset was first employed to evaluate the effectiveness of the proposed approach. This dataset represents one of the most widely studied binary classification benchmarks in the healthcare domain. The performance of the proposed CAQFM variants was compared with conventional quantum feature maps.

\begin{table}[H]
\centering
\caption{Results obtained on the Breast Cancer dataset}
\label{tab:bc_results}

\setlength{\tabcolsep}{3pt}
\small

\begin{tabular}{lccccccc}
\hline
\textbf{Method} &
\textbf{Accuracy} &
\textbf{Col. Gates} &
\textbf{Precision} &
\textbf{Recall} &
\textbf{F1} &
\textbf{ROC-AUC} &
\textbf{Epoch} \\
\hline
Pearson-CAQFM   & 0.94 & 1  & 0.92 & 0.92 & 0.92 & 0.98 & 31 \\
Spearman-CAQFM & 0.95 & 3  & 0.97 & 0.88 & 0.93 & 0.99 & 19 \\
Kendall-CAQFM  & 0.95 & 1  & 0.96 & 0.90 & 0.93 & 0.99 & 56 \\
MI-CAQFM       & 0.93 & 1  & 0.97 & 0.83 & 0.89 & 0.97 & 19 \\
Distance-CAQFM & 0.91 & 2  & 0.84 & 0.94 & 0.89 & 0.93 & 25 \\
Z Feature Map  & 0.87 & 0  & 0.92 & 0.71 & 0.80 & 0.94 & 74 \\
ZZ Feature Map & 0.85 & 6  & 0.84 & 0.73 & 0.78 & 0.94 & 93 \\
Pauli Feature Map & 0.86 & 36 & 0.94 & 0.66 & 0.77 & 0.93 & 92 \\
\hline
\end{tabular}
\end{table}

For all experiments on the Breast Cancer dataset, the dependency threshold was set to 0.95, the variational ansatz consisted of a single layer, and the classification threshold was fixed at 0.4. The corresponding training and test performance curves are presented in Figures 1,2,3,4.

\begin{figure}[!ht]
\centering
\includegraphics[width=0.90\textwidth]{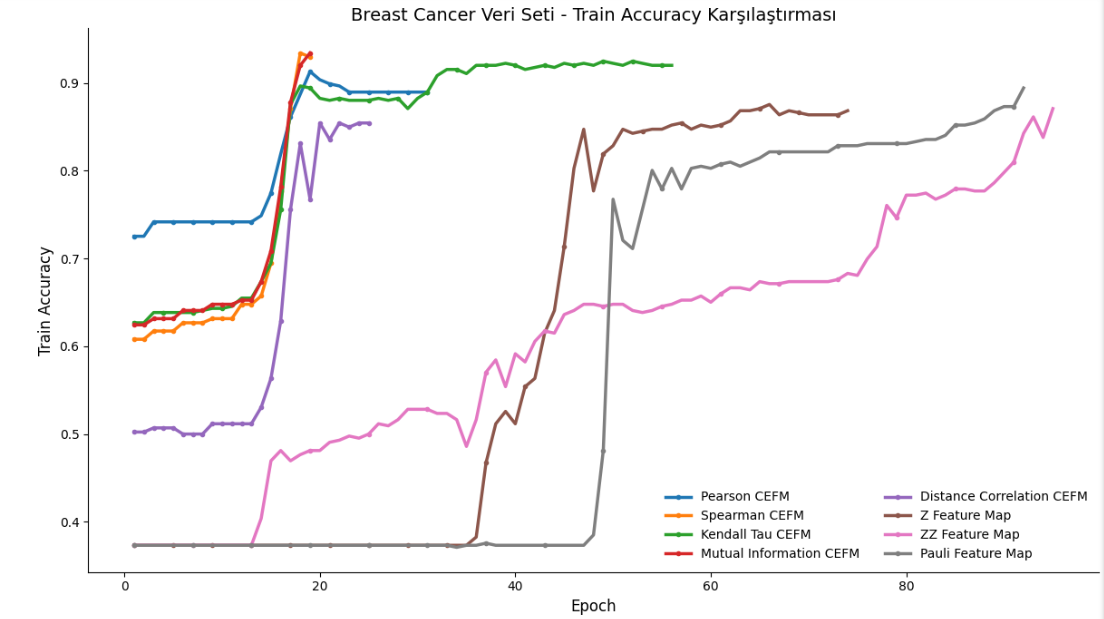}
\caption{Training accuracy values throughout the epochs for the Breast Cancer dataset.}
\label{fig:breast_train_acc}
\end{figure}

\begin{figure}[!ht]
\centering
\includegraphics[width=0.90\textwidth]{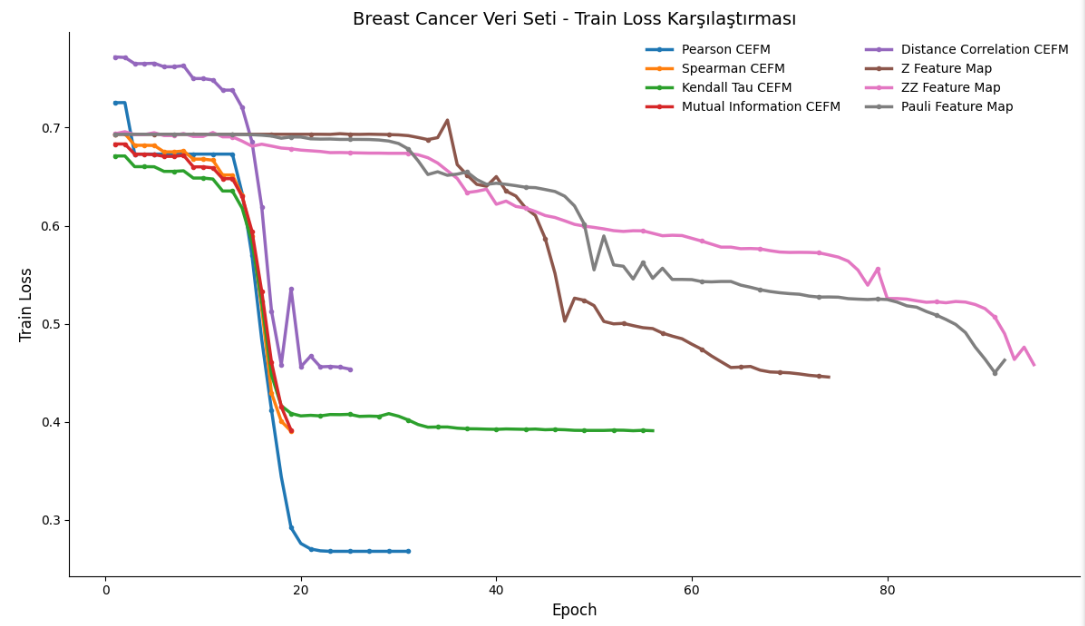}
\caption{Training loss values throughout the epochs for the Breast Cancer dataset.}
\label{fig:breast_train_loss}
\end{figure}

\begin{figure}[!ht]
\centering
\includegraphics[width=0.90\textwidth]{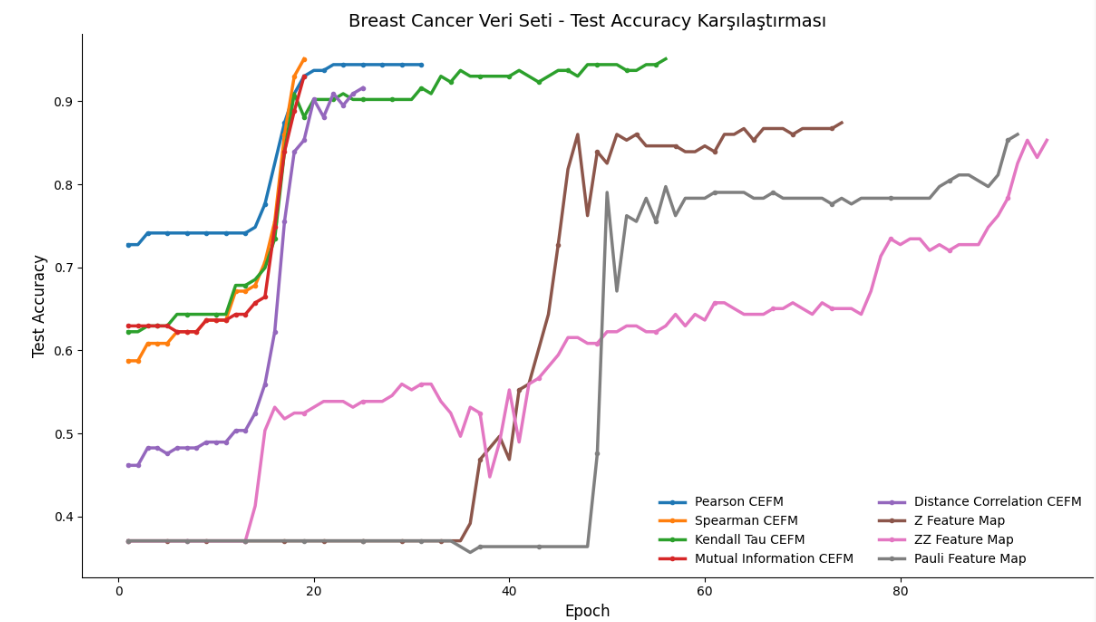}
\caption{Test accuracy values throughout the epochs for the Breast Cancer dataset.}
\label{fig:breast_test_acc}
\end{figure}

\begin{figure}[!ht]
\centering
\includegraphics[width=0.90\textwidth]{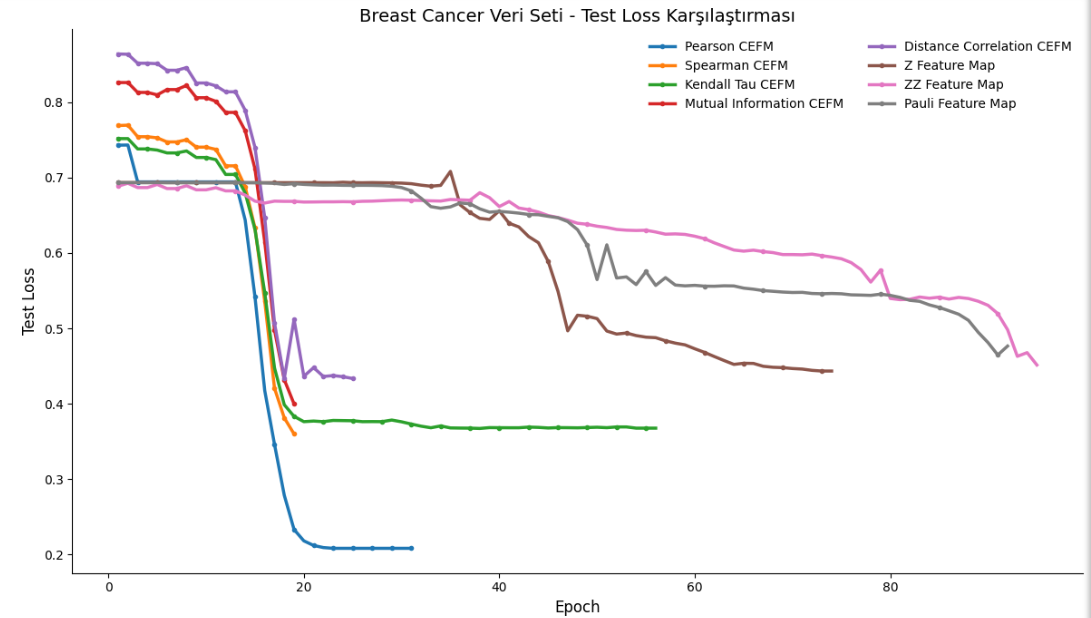}
\caption{Test loss values throughout the epochs for the Breast Cancer dataset.}
\label{fig:breast_test_loss}
\end{figure}
```

The results indicate that the proposed CAQFM-based approaches consistently outperform conventional quantum feature maps. Among the dependency-aware methods, Spearman-CAQFM and Kendall-CAQFM achieved the highest classification performance with an accuracy of approximately \(95.1\%\). Pearson-CAQFM followed closely with an accuracy of \(94.4\%\), while MI-CAQFM and Distance-CAQFM reached \(93.0\%\) and \(91.6\%\), respectively. These findings suggest that incorporating statistical dependencies into the quantum encoding process enables the quantum circuit to capture more discriminative information from the dataset.

In contrast, the conventional feature maps produced lower classification performance. The best accuracy achieved by the Z Feature Map was approximately \(87.4\%\), while the ZZ Feature Map and Pauli Feature Map reached \(85.3\%\) and \(86.0\%\), respectively. Although these results remain competitive, they consistently lag behind the CAQFM variants.

A similar trend can be observed in the loss values. Pearson-CAQFM achieved the lowest test loss of approximately 0.2084, indicating not only high classification accuracy but also more reliable class probability estimates. The minimum test losses obtained by Spearman-CAQFM, Kendall-CAQFM, MI-CAQFM, and Distance-CAQFM were approximately 0.3600, 0.3677, 0.3998, and 0.4335, respectively. In comparison, the minimum test losses of the Z, ZZ, and Pauli Feature Maps were approximately 0.4434, 0.4514, and 0.4647.

From a convergence perspective, the advantages of CAQFM are also evident. Spearman-CAQFM, Kendall-CAQFM, and MI-CAQFM reached high accuracy levels within approximately 17--19 epochs, whereas the conventional feature maps required substantially more training iterations. In particular, the Z Feature Map exhibited noticeable improvements only after approximately 40 epochs, while the ZZ and Pauli Feature Maps remained in lower-accuracy regions for a longer period.

Overall, the experimental results demonstrate that incorporating statistical dependencies into the quantum encoding process can significantly improve classification performance. Compared with conventional Z, ZZ, and Pauli Feature Maps, the proposed CAQFM variants consistently achieved higher accuracy, lower loss values, and faster convergence. Among the investigated approaches, Pearson-, Spearman-, and Kendall-based CAQFM models exhibited the most competitive performance, indicating that dependency-aware quantum representations can provide a more informative feature space for variational quantum classifiers.

\subsection{Credit Default Dataset Results}

The second set of experiments was conducted on the Credit Default dataset. This dataset represents a financial classification problem characterized by complex decision boundaries and potentially nonlinear relationships among variables. The performance of the proposed CAQFM variants was compared against conventional quantum feature maps.

\begin{table}[H]
\centering
\caption{Results obtained on the Credit Default dataset}
\label{tab:credit_results}

\setlength{\tabcolsep}{3pt}
\small

\begin{tabular}{lccccccc}
\hline
\textbf{Method} &
\textbf{Accuracy} &
\textbf{CRY Gates} &
\textbf{Precision} &
\textbf{Recall} &
\textbf{F1} &
\textbf{ROC-AUC} &
\textbf{Epoch} \\
\hline
Pearson-CAQFM      & 0.80 & 2  & 0.71 & 1.00 & 0.83 & 0.87 & 55 \\
Spearman-CAQFM     & 0.81 & 2  & 0.71 & 1.00 & 0.83 & 0.87 & 64 \\
Kendall-CAQFM      & 0.81 & 1  & 0.71 & 1.00 & 0.83 & 0.87 & 34 \\
MI-CAQFM           & 0.81 & 1  & 0.73 & 1.00 & 0.84 & 0.85 & 30 \\
Distance-CAQFM     & 0.80 & 2  & 0.70 & 1.00 & 0.82 & 0.87 & 50 \\
Z Feature Map      & 0.74 & 0  & 0.70 & 0.85 & 0.77 & 0.82 & 28 \\
ZZ Feature Map     & 0.80 & 6  & 0.74 & 0.92 & 0.82 & 0.87 & 88 \\
Pauli Feature Map  & 0.70 & 18 & 0.65 & 0.81 & 0.72 & 0.77 & 70 \\
\hline
\end{tabular}

\end{table}

For all experiments on the Credit Default dataset, the dependency threshold was set to 0.20, the variational ansatz consisted of a single layer, and the classification threshold was fixed at 0.5. The corresponding training and test performance curves are presented in Figures 5,6,7 and 8.

\begin{figure}[!ht]
\centering
\includegraphics[width=0.90\textwidth]{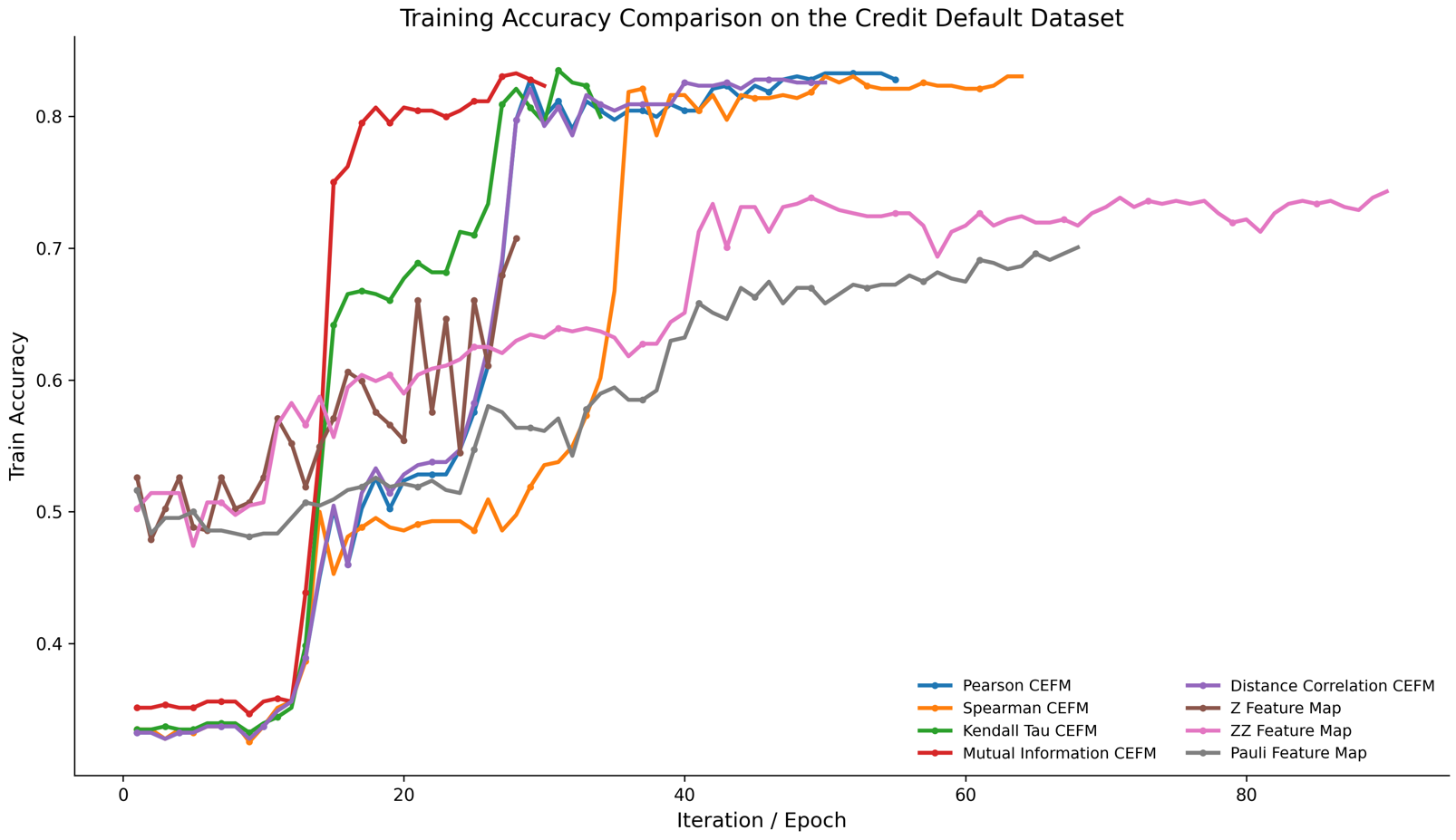}
\caption{Training accuracy values throughout the epochs for the Credit Default dataset.}
\label{fig:credit_train_acc}
\end{figure}

\begin{figure}[!ht]
\centering
\includegraphics[width=0.90\textwidth]{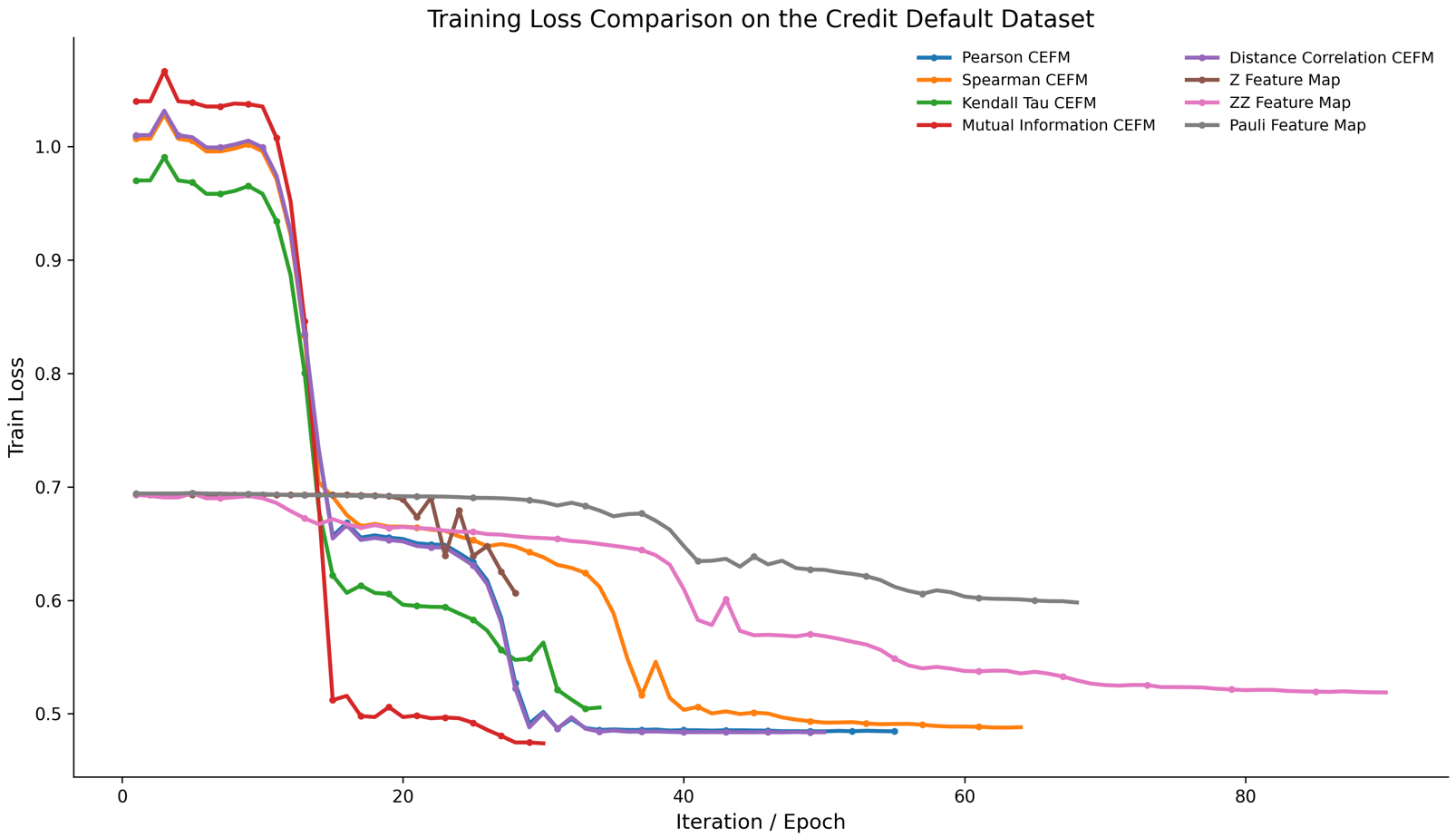}
\caption{Training loss values throughout the epochs for the Credit Default dataset.}
\label{fig:credit_train_loss}
\end{figure}

\begin{figure}[!ht]
\centering
\includegraphics[width=0.90\textwidth]{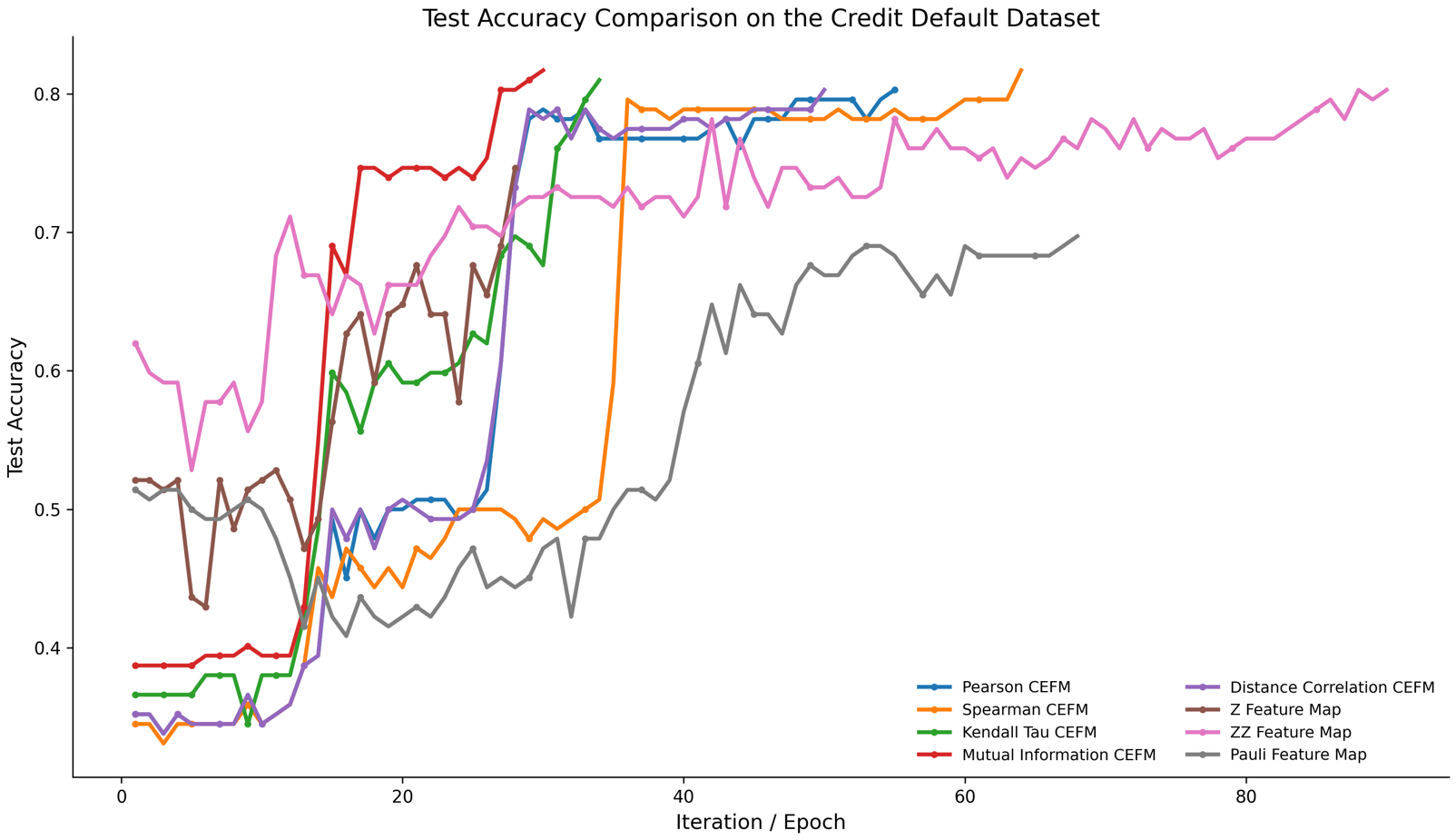}
\caption{Test accuracy values throughout the epochs for the Credit Default dataset.}
\label{fig:credit_test_acc}
\end{figure}

\begin{figure}[!ht]
\centering
\includegraphics[width=0.90\textwidth]{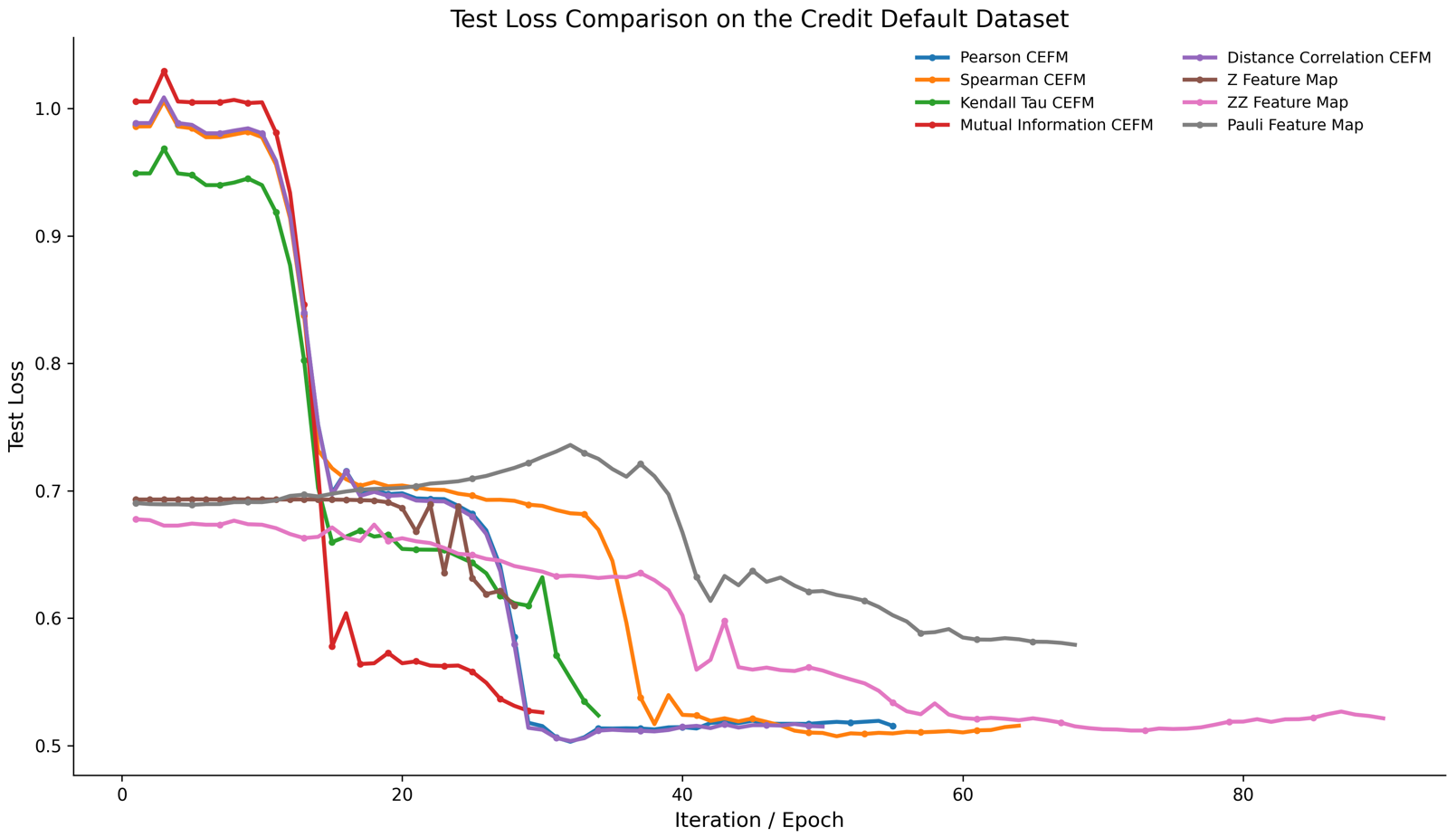}
\caption{Test loss values throughout the epochs for the Credit Default dataset.}
\label{fig:credit_test_loss}
\end{figure}

The results demonstrate that dependency-aware quantum feature maps provide a consistent advantage over conventional quantum encoding techniques. Among all evaluated approaches, Spearman-CAQFM and MI-CAQFM achieved the highest classification accuracy of approximately \(81.7\%\). Kendall-CAQFM followed closely with an accuracy of \(81.0\%\), while Pearson-CAQFM and Distance-CAQFM achieved approximately \(80.3\%\). These results indicate that incorporating feature dependencies into the quantum encoding process can improve the representation of financial data characterized by complex variable interactions.

Among the conventional feature maps, ZZ Feature Map achieved the strongest performance with an accuracy comparable to Pearson-CAQFM and Distance-CAQFM. However, ZZ Feature Map required six predefined entangling interactions and considerably more training epochs to converge. In contrast, the CAQFM variants achieved similar or superior performance using significantly fewer data-driven interactions. This observation suggests that dependency-guided quantum connectivity can provide a more efficient encoding strategy than fixed entanglement topologies.

The ranking-based dependency measures, namely Spearman Correlation and Kendall Tau, exhibited particularly strong performance. Since credit default prediction often involves monotonic but not strictly linear relationships among variables, rank-based dependency measures appear to capture the underlying structure of the dataset more effectively than purely linear correlation measures. This observation is supported by the superior performance of Spearman-CAQFM and the faster convergence of Kendall-CAQFM, which reached its final performance level after only 34 epochs.

Mutual Information-based CAQFM produced one of the most competitive results while requiring only a single controlled interaction and 30 training epochs. Given that Mutual Information is capable of capturing both linear and nonlinear dependencies, its strong performance suggests that information-theoretic relationships play an important role in the Credit Default dataset. The rapid convergence observed for this approach further indicates that incorporating informative dependency structures into the quantum encoding stage can facilitate optimization.

In contrast, the Z Feature Map and Pauli Feature Map achieved noticeably lower performance. The Z Feature Map reached an accuracy of approximately \(74.7\%\), while the Pauli Feature Map remained near \(70\%\). These results suggest that encoding schemes that do not explicitly exploit dataset-specific dependency structures may be less effective for complex financial classification problems.

Overall, the Credit Default experiments confirm the benefits of incorporating statistical dependency information into quantum feature maps. Compared with conventional quantum encoding strategies, the proposed CAQFM variants achieved higher classification accuracy, required fewer meaningful interactions, and exhibited faster convergence behavior. The strongest results were obtained using Spearman Correlation, Kendall Tau, and Mutual Information, indicating that dependency-aware quantum representations can effectively capture the complex relationships present in financial datasets.

\subsection{Student Placement Dataset Results}

The third set of experiments was conducted on the Student Placement dataset, which represents a binary classification problem in the educational domain. The objective of this dataset is to predict student placement outcomes based on academic performance and related attributes.

\begin{table}[H]
\centering
\caption{Results obtained on the Student Placement dataset}
\label{tab:student_results}

\setlength{\tabcolsep}{3pt}
\small

\begin{tabular}{lccccccc}
\hline
\textbf{Method} &
\textbf{Accuracy} &
\textbf{CRY Gates} &
\textbf{Precision} &
\textbf{Recall} &
\textbf{F1} &
\textbf{ROC-AUC} &
\textbf{Epoch} \\
\hline
Pearson-CAQFM      & 0.97 & 4  & 1.00 & 0.93 & 0.96 & 1.00 & 39 \\
Spearman-CAQFM     & 0.97 & 4  & 1.00 & 0.93 & 0.96 & 1.00 & 39 \\
Kendall-CAQFM      & 0.98 & 3  & 1.00 & 0.97 & 0.98 & 1.00 & 18 \\
MI-CAQFM           & 0.99 & 3  & 1.00 & 0.98 & 0.98 & 1.00 & 18 \\
Distance-CAQFM     & 0.98 & 3  & 1.00 & 0.97 & 0.98 & 1.00 & 18 \\
Z Feature Map      & 0.99 & 0  & 1.00 & 0.98 & 0.99 & 1.00 & 23 \\
ZZ Feature Map     & 0.70 & 12 & 0.82 & 0.48 & 0.61 & 0.82 & 47 \\
Pauli Feature Map  & 0.75 & 36 & 0.73 & 0.76 & 0.75 & 0.79 & 21 \\
\hline
\end{tabular}

\end{table}

For all experiments on the Student Placement dataset, the dependency threshold was set to 0.10, the variational ansatz consisted of a single layer, and the classification threshold was fixed at 0.5. The corresponding training and test performance curves are presented in Figures 9,10,11 and 12.

\begin{figure}[!ht]
\centering
\includegraphics[width=0.90\textwidth]{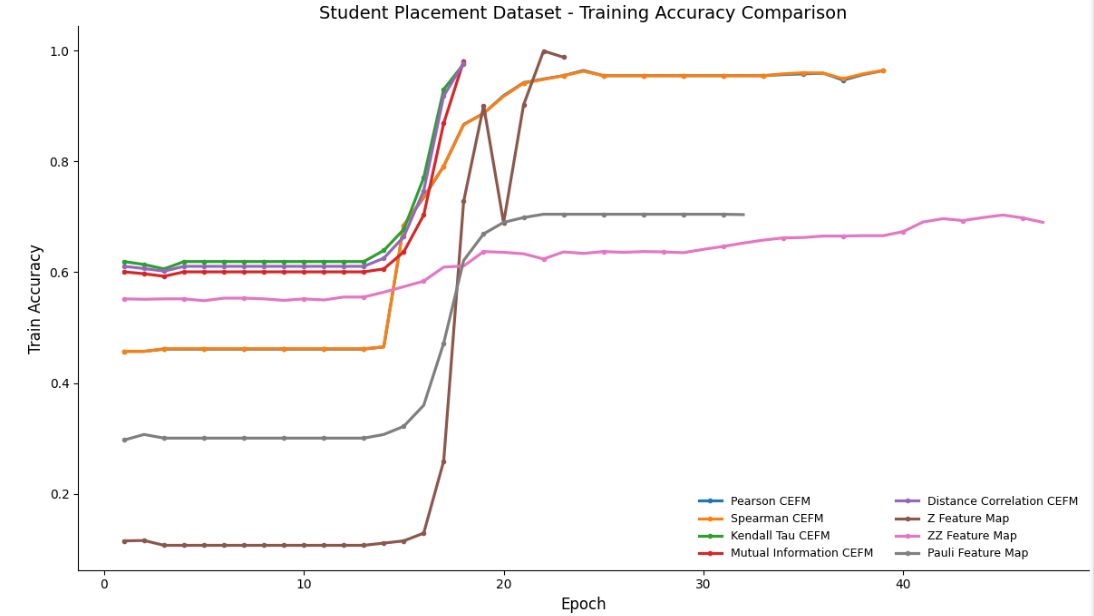}
\caption{Training accuracy values throughout the epochs for the Student Placement dataset.}
\label{fig:student_train_acc}
\end{figure}

\begin{figure}[!ht]
\centering
\includegraphics[width=0.90\textwidth]{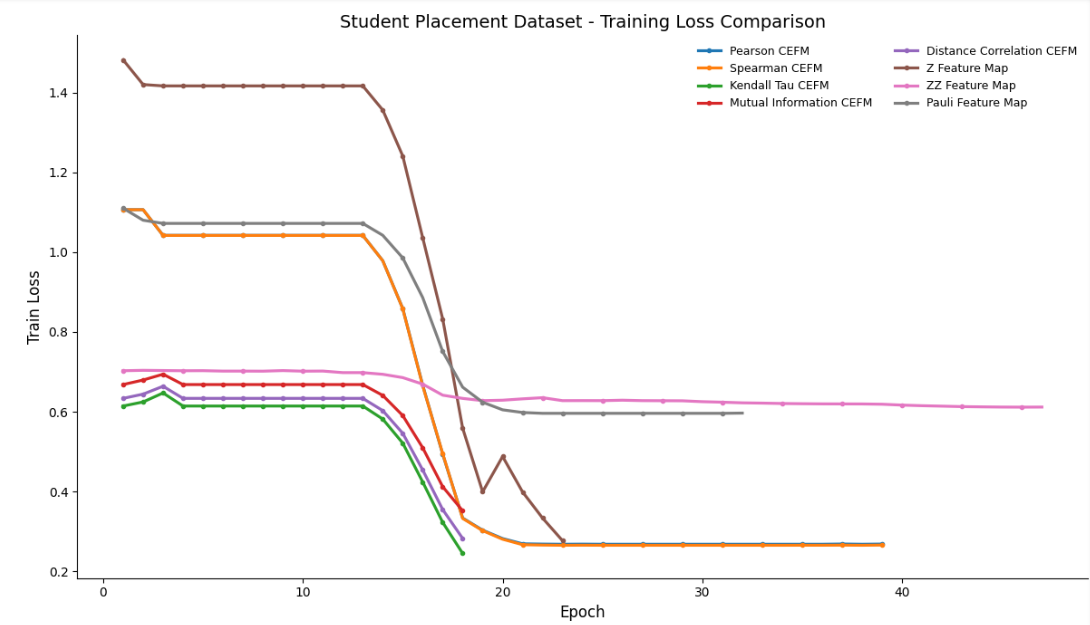}
\caption{Training loss values throughout the epochs for the Student Placement dataset.}
\label{fig:student_train_loss}
\end{figure}

\begin{figure}[!ht]
\centering
\includegraphics[width=0.90\textwidth]{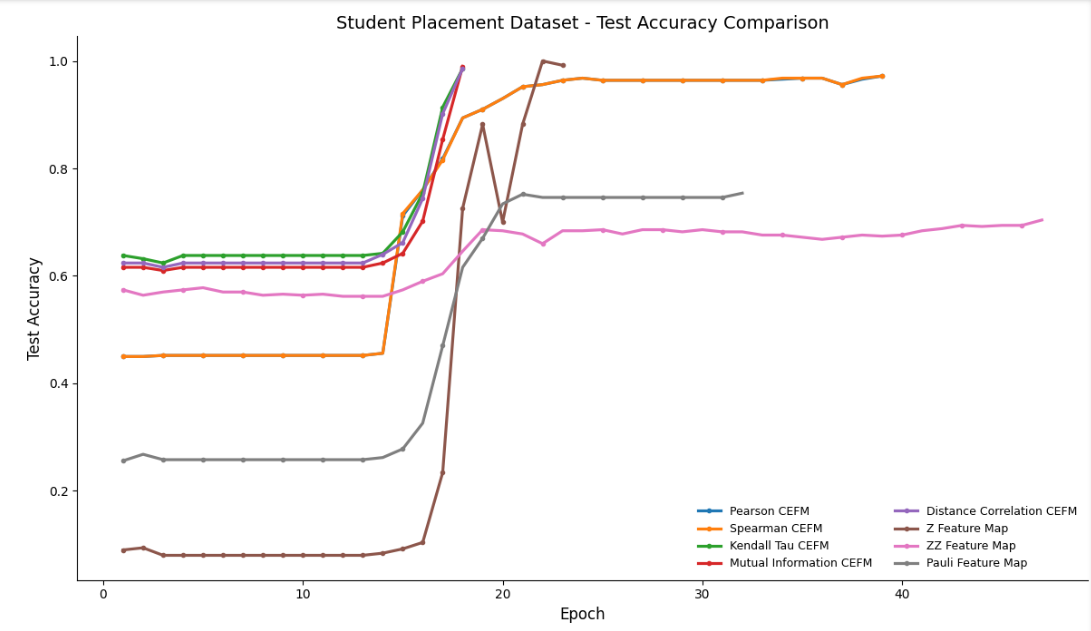}
\caption{Test accuracy values throughout the epochs for the Student Placement dataset.}
\label{fig:student_test_acc}
\end{figure}

\begin{figure}[!ht]
\centering
\includegraphics[width=0.90\textwidth]{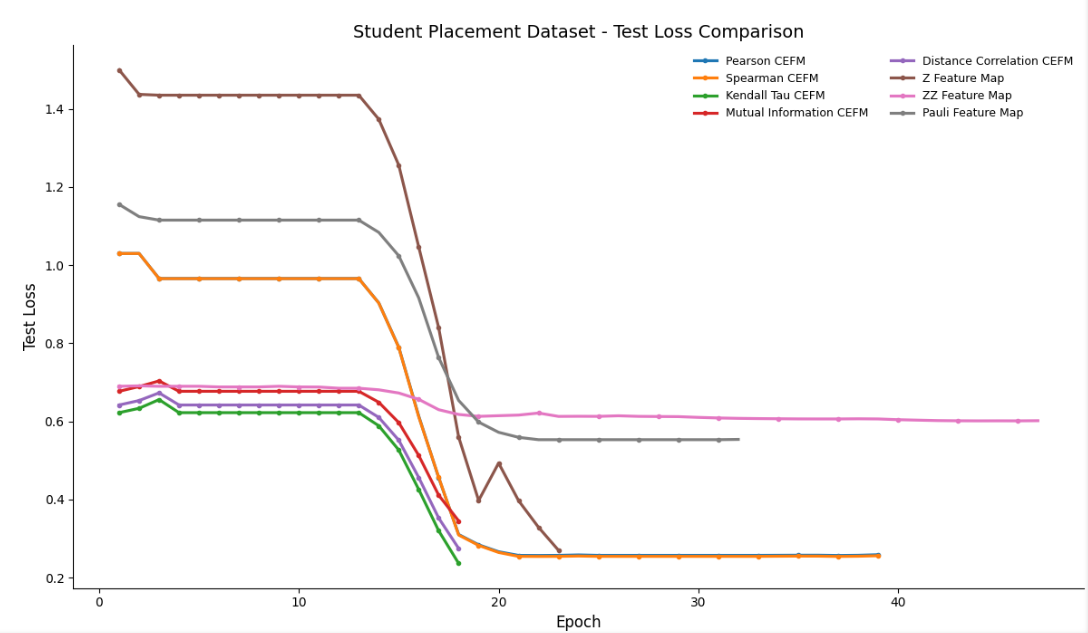}
\caption{Test loss values throughout the epochs for the Student Placement dataset.}
\label{fig:student_test_loss}
\end{figure}

The Student Placement dataset produced the highest classification performance among all datasets considered in this study. Most CAQFM variants achieved accuracies exceeding \(97\%\), indicating that the selected features contain highly discriminative information for predicting placement outcomes.

Among the dependency-aware approaches, MI-CAQFM achieved the strongest performance with an accuracy of approximately \(99\%\), followed by Kendall-CAQFM and Distance-CAQFM with accuracies close to \(98.6\%\). Pearson-CAQFM and Spearman-CAQFM also produced highly competitive results, reaching approximately \(97.2\%\) accuracy. These findings suggest that dependency-aware quantum representations are highly effective for educational decision-support problems, where multiple academic factors jointly influence the target outcome.

The ranking-based dependency measures, namely Spearman Correlation and Kendall Tau, again demonstrated strong performance. However, unlike the Credit Default dataset, the performance differences among the CAQFM variants were relatively small. This observation indicates that the predictive information in the Student Placement dataset is sufficiently strong to be captured by several different dependency measures.

A notable result is the performance of the Z Feature Map, which achieved approximately \(99.2\%\) accuracy and slightly outperformed all CAQFM variants. While this result appears highly favorable, it should be interpreted with caution. The absence of entangling interactions in the Z Feature Map suggests that the dataset may be largely separable using individual feature information alone. In other words, the dominant predictive patterns may already be encoded within single features, reducing the need for complex feature interactions.

In contrast, the ZZ Feature Map and Pauli Feature Map produced substantially lower classification performance. The ZZ Feature Map achieved approximately \(70.4\%\) accuracy, while the Pauli Feature Map reached approximately \(75.4\%\). These results indicate that predefined entanglement structures do not necessarily provide an advantage when the underlying dataset can be effectively represented using simpler feature relationships.

From a convergence perspective, MI-CAQFM, Kendall-CAQFM, and Distance-CAQFM exhibited the fastest learning behavior, reaching their highest performance levels within only 18 epochs. Pearson-CAQFM and Spearman-CAQFM required longer training periods, converging after approximately 39 epochs. Nevertheless, all CAQFM variants demonstrated stable optimization behavior and consistently achieved high classification performance.

Overall, the results obtained on the Student Placement dataset further support the effectiveness of incorporating statistical dependency information into quantum feature maps. Although the Z Feature Map achieved the highest accuracy in this particular dataset, the CAQFM variants consistently provided competitive performance while maintaining an interpretable connection between dataset statistics and quantum circuit structure. The strong results obtained by MI-, Kendall-, and Distance-based CAQFM models indicate that dependency-aware quantum representations can serve as a powerful approach for educational prediction and decision-support tasks.

\subsection{Discussion}

The experimental results demonstrate that incorporating statistical dependency information into the quantum encoding process can improve classification performance across multiple application domains. Unlike conventional quantum feature maps, which rely on fixed interaction topologies independent of the underlying dataset, the proposed CAQFM framework adapts the quantum circuit structure according to the statistical relationships observed among features.

The results obtained from the Breast Cancer dataset indicate that dependency-aware quantum representations can significantly enhance classification performance. In particular, Pearson-, Spearman-, and Kendall-based CAQFM variants consistently outperformed conventional Z, ZZ, and Pauli Feature Maps in terms of accuracy, convergence speed, and loss values. These findings suggest that exploiting feature dependencies during quantum encoding enables the quantum circuit to capture more discriminative information from biomedical datasets.

For the Credit Default dataset, the strongest performance was achieved by Spearman-, Kendall-, and Mutual Information-based CAQFM variants. The superiority of these methods indicates that rank-based and information-theoretic dependency measures are particularly effective for modeling the complex relationships commonly encountered in financial datasets. Furthermore, CAQFM approaches achieved comparable or superior performance using fewer meaningful interactions than conventional feature maps with fixed entanglement structures.

The Student Placement dataset exhibited a different behavior. While CAQFM variants achieved highly competitive accuracies exceeding \(97\%\), the Z Feature Map obtained the highest overall accuracy. This observation suggests that the dataset may already possess highly informative individual features, reducing the necessity for complex interaction structures. Nevertheless, CAQFM variants maintained strong performance while providing a more interpretable connection between dataset statistics and quantum circuit architecture.

Another notable observation concerns the relationship between dependency measures and dataset characteristics. Pearson Correlation generally performed well when linear relationships were dominant, whereas Spearman Correlation and Kendall Tau demonstrated advantages when monotonic dependencies were more informative. Mutual Information and Distance Correlation, which are capable of capturing nonlinear relationships, proved particularly effective in datasets containing more complex dependency structures. These findings indicate that the selection of an appropriate dependency measure plays a crucial role in the effectiveness of dependency-aware quantum feature maps.

Overall, the proposed CAQFM framework provides a flexible and data-driven quantum encoding mechanism capable of adapting to the statistical characteristics of different datasets. The results suggest that dependency-aware quantum representations constitute a promising direction for the development of next-generation quantum machine learning models.

\section{Conclusion}

In this study, a Correlation-Aware Quantum Feature Map (CAQFM) was proposed to incorporate statistical dependencies among features directly into the quantum data encoding process. Unlike conventional quantum feature maps that employ predefined interaction structures, the proposed approach constructs quantum interactions according to the statistical properties of the dataset, thereby introducing a data-driven mechanism for quantum representation learning.

The proposed framework utilizes Pearson Correlation, Spearman Correlation, Kendall Tau, Mutual Information, and Distance Correlation to quantify pairwise feature dependencies. The resulting dependency coefficients are subsequently transformed into controlled quantum rotations, allowing relational information present in the dataset to be embedded into the quantum feature map. As a result, the quantum representation captures not only individual feature values but also the statistical relationships among features.

The effectiveness of the proposed approach was evaluated on three binary classification problems from healthcare, finance, and education domains, namely Breast Cancer diagnosis, Credit Default prediction, and Student Placement prediction. Experimental results demonstrated that CAQFM-based models consistently achieved competitive and, in many cases, superior performance compared with conventional Z, ZZ, and Pauli Feature Maps. In addition to improved classification accuracy, the proposed framework frequently exhibited faster convergence and required fewer meaningful interactions to achieve strong predictive performance.

One of the primary contributions of this work is the establishment of an interpretable connection between classical dependency analysis and quantum data encoding. The proposed framework demonstrates that the design of quantum feature maps can be guided not only by quantum circuit considerations but also by the statistical structure of the underlying dataset. This perspective introduces a new direction for developing adaptive and data-aware quantum representations in quantum machine learning.

Future work will focus on evaluating the proposed approach on real quantum hardware, developing adaptive threshold-selection strategies, and extending dependency-aware principles to the design of variational ansatz architectures. In addition, the application of CAQFM to multiclass classification problems, larger datasets, and graph-based quantum representations constitutes a promising avenue for further research.

\section*{Data Availability}

All datasets used in this study are publicly available and can be accessed through the Kaggle platform.

\begin{itemize}
\item Breast Cancer Wisconsin Diagnostic Dataset:\\
\url{https://www.kaggle.com/datasets/utkarshx27/breast-cancer-wisconsin-diagnostic-dataset}

\item Default of Credit Card Clients Dataset:\\
\url{https://www.kaggle.com/datasets/uciml/default-of-credit-card-clients-dataset}

\item Student Placement Dataset:\\
\url{https://www.kaggle.com/datasets/ruchikakumbhar/student-placement-dataset}
\end{itemize}

No proprietary or restricted-access data were used in this study. All datasets were accessed between January and May 2026 and were used in accordance with their respective licenses and terms of use.


\begin{thebibliography}{99}

\bibitem{benioff1980}
P. Benioff,
``The Computer as a Physical System: A Microscopic Quantum Mechanical Hamiltonian Model of Computers as Represented by Turing Machines,''
\textit{Journal of Statistical Physics},
22(5), 563--591, 1980.

\bibitem{feynman1982}
R. P. Feynman,
``Simulating Physics with Computers,''
\textit{International Journal of Theoretical Physics},
21(6--7), 467--488, 1982.

\bibitem{deutsch1985}
D. Deutsch,
``Quantum Theory, the Church--Turing Principle and the Universal Quantum Computer,''
\textit{Proceedings of the Royal Society of London A},
400(1818), 97--117, 1985.

\bibitem{bennett1995}
C. H. Bennett,
``Quantum Information and Computation,''
\textit{Physics Today},
48(10), 24--30, 1995.

\bibitem{preskill2018}
J. Preskill,
``Quantum Computing in the NISQ Era and Beyond,''
\textit{Quantum},
2, 79, 2018.

\bibitem{AbuGhanem2023}
M. AbuGhanem and H. Eleuch,
``NISQ Computers: A Path to Quantum Supremacy,''
\textit{IEEE Access},
12, 102941--102961, 2023.

\bibitem{lohia2024}
A. Lohia,
``Quantum Artificial Intelligence: Enhancing Machine Learning with Quantum Computing,''
\textit{Journal of Quantum Science and Technology},
1(2), 6--15, 2024.

\bibitem{klusch2024}
M. Klusch, J. Laessig, D. Muessig, A. Macaluso and F. K. Wilhelm,
``Quantum Artificial Intelligence: A Brief Survey,''
\textit{KI -- Kuenstliche Intelligenz},
38(4), 509--520, 2024.

\bibitem{acampora2025}
G. Acampora, A. Chiatto, R. Schiattarella and A. Vitiello,
``Quantum Artificial Intelligence: A Survey,''
\textit{Computer Science Review},
59, 100807, 2025.

\bibitem{Sharma2024}
S. Sharma and R. N.,
``Survey of Encoding Techniques for Quantum Machine Learning,''
\textit{Cybernetics and Physics},
13(2), 2024.

\bibitem{rath2024}
M. Rath and H. Date,
``Quantum Data Encoding: A Comparative Analysis of Classical-to-Quantum Mapping Techniques and Their Impact on Machine Learning Accuracy,''
\textit{EPJ Quantum Technology},
vol. 11, no. 1, p. 72, 2024.

\bibitem{khan}
M. A. Khan, M. N. Aman, and B. Sikdar,
``Beyond Bits: A Review of Quantum Embedding Techniques for Efficient Information Processing,''
\textit{IEEE Access},
vol. 12, pp. 46118--46137, 2024.
doi: 10.1109/ACCESS.2024.3382150.

\bibitem{weigold}
M. Weigold, J. Barzen, F. Leymann, and M. Salm,
``Data Encoding Patterns for Quantum Computing,''
in \textit{Proceedings of the Pattern Languages of Programs Conference (PLoP)},
2022, Article 2.

\bibitem{Singh2025}
N. Singh and S. R. Pokhrel,
``Modeling Feature Maps for Quantum Machine Learning,''
\textit{arXiv preprint arXiv:2501.08205},
2025.

\bibitem{suzuki2024}
T. Suzuki, T. Hasebe, and T. Miyazaki,
``Quantum Support Vector Machines for Classification and Regression on a Trapped-Ion Quantum Computer,''
\textit{Quantum Machine Intelligence},
vol. 6, pp. 1--14, 2024.
doi: 10.1007/s42484-024-00165-0.

\bibitem{ozpolat2024}
Z. Ozpolat, O. Yildirim, and M. Karabatak,
``The Effect of Linear Discriminant Analysis and Quantum Feature Maps on QSVM Performance for Obesity Diagnosis,''
\textit{Balkan Journal of Electrical and Computer Engineering},
2024.
doi: 10.17694/bajece.1475896.

\bibitem{bartkiewicz2020}
K. Bartkiewicz, C. Gneiting, A. Cernoch, K. Jirakova, K. Lemr, and F. Nori,
``Experimental Kernel-Based Quantum Machine Learning in Finite Feature Space,''
\textit{Scientific Reports},
vol. 10, no. 1, p. 12356, 2020.
doi: 10.1038/s41598-020-68911-5.

\bibitem{toufah2025}
A. Toufah, M. A. Kadim, and M. Y. El Hafidi,
``Investigating Quantum Feature Maps in Quantum Support Vector Machines for Lung Cancer Classification,''
\textit{Journal of Artificial Intelligence Research and Innovation},
vol. 1, no. 1, pp. 100--110, 2025.
doi: 10.29328/journal.jairi.1001012.

\bibitem{Alami2024}
M. El Alami, N. Innan, M. Shafique, and M. Bennai,
``Comparative Performance Analysis of Quantum Machine Learning Architectures for Credit Card Fraud Detection,''
\textit{Applied Intelligence},
vol. 56, p. 83, 2026.
doi: 10.1007/s10489-026-07110-7.

\bibitem{Hirai2024}
H. Hirai,
``Practical Application of Quantum Neural Network to Materials Informatics,''
\textit{Scientific Reports},
vol. 14, p. 8583, 2024.
doi: 10.1038/s41598-024-59276-0.

\bibitem{ArthurDate2022}
D. Arthur and P. Date,
``A Hybrid Quantum-Classical Neural Network Architecture for Binary Classification,''
\textit{arXiv preprint arXiv:2201.01820},
2022.

\bibitem{Tyrovolas2024}
K. A. Tychola, T. Kalampokas, and G. A. Papakostas,
``Quantum Machine Learning---An Overview,''
\textit{Electronics},
vol. 13, no. 3, p. 560, 2024.
doi: 10.3390/electronics13030560.

\bibitem{Idzikowski2026}
R. Idzikowski, M. A. Kucharski, K. Pempera, and M. Jaroszczuk,
``A Survey on Quantum Machine Learning Applications in Medicine and Healthcare,''
\textit{Applied Sciences},
vol. 16, no. 3, p. 1630, 2026.
doi: 10.3390/app16031630.

\bibitem{MurrayParsons2024}
S. A. Murray and E. Parsons,
``Advancing EEG Signal Analysis with Quantum Machine Learning,''
\textit{arXiv preprint arXiv:2409.11462},
2024.

\bibitem{HowCheah2024}
M.-L. How and S.-M. Cheah,
``Forging the Future: Strategic Approaches to Quantum AI Integration for Industry Transformation,''
\textit{AI},
vol. 5, no. 1, pp. 290--323, 2024.
doi: 10.3390/ai5010015.

\bibitem{BuresFrantis2024}
M. Bures and P. Frantis,
``Future Use of Quantum Artificial Intelligence in Military Domains,''
in \textit{2024 International Conference on Military Technologies (ICMT)},
pp. 1--6, 2024.
doi: 10.1109/ICMT60383.2024.10579948.

\bibitem{Spearman1904}
C. Spearman,
``The Proof and Measurement of Association Between Two Things,''
\textit{The American Journal of Psychology},
vol. 15, no. 1, pp. 72--101, 1904.
doi: 10.2307/1412159.

\bibitem{deWinter2016}
J. C. F. de Winter, S. D. Gosling, and J. Potter,
``Comparing the Pearson and Spearman Correlation Coefficients Across Distributions and Sample Sizes: A Tutorial Using Simulations and Empirical Data,''
\textit{Psychological Methods},
vol. 21, no. 3, pp. 273--290, 2016.

\bibitem{Stepanov2015O}
A. Stepanov,
``On the Kendall Correlation Coefficient,''
\textit{arXiv preprint},
2015.

\bibitem{TisocBeltran2022}
M. Tisoc and J. V. Beltran,
``Mutual Information: A Way to Quantify Correlations,''
\textit{Revista Brasileira de Ensino de Física},
vol. 44, p. e20220055, 2022.
doi: 10.1590/1806-9126-RBEF-2022-0055.



\bibitem{Edelmann2021}
D. Edelmann, T. F. Mori, and G. J. Szekely,
``On Relationships Between the Pearson and the Distance Correlation Coefficients,''
\textit{Statistics  Probability Letters},
vol. 169, p. 108960, 2021.
doi: 10.1016/j.spl.2020.108960.

\bibitem{Cerezo2021}
M. Cerezo, A. Arrasmith, R. Babbush, S. C. Benjamin,
S. Endo, K. Fujii, J. R. McClean, K. Mitarai,
X. Yuan, L. Cincio, and P. J. Coles,
``Variational Quantum Algorithms,''
\textit{Nature Reviews Physics},
vol. 3, no. 9, pp. 625--644, 2021.
doi: 10.1038/s42254-021-00348-9.

\bibitem{Qi2024}
H. Qi, S. Xiao, Z. Liu, C. Gong, and A. Gani,
``Variational Quantum Algorithms: Fundamental Concepts, Applications and Challenges,''
\textit{Quantum Information Processing},
vol. 23, no. 6, p. 224, 2024.
doi: 10.1007/s11128-024-04438-2.

\bibitem{Endo2020}
S. Endo, J. Sun, Y. Li, S. C. Benjamin, and X. Yuan,
``Variational Quantum Simulation of General Processes,''
\textit{Physical Review Letters},
vol. 125, no. 1, p. 010501, 2020.
doi: 10.1103/PhysRevLett.125.010501.

\bibitem{Wu2021}
A. Wu, G. Li, Y. Wang, B. Feng, Y. Ding, and Y. Xie,
``Towards Efficient Ansatz Architecture for Variational Quantum Algorithms,''
\textit{arXiv preprint arXiv:2111.13730},
2021.

\bibitem{Qin2023}
J. Qin,
``Review of Ansatz Designing Techniques for Variational Quantum Algorithms,''
\textit{Journal of Physics: Conference Series},
vol. 2634, no. 1, p. 012043, 2023.
doi: 10.1088/1742-6596/2634/1/012043.

\bibitem{danyal}
D. Maheshwari, D. Sierra-Sosa, and B. Garcia-Zapirain,
``Variational Quantum Classifier for Binary Classification: Real vs Synthetic Dataset,''
\textit{IEEE Access},
vol. 10, pp. 3705--3715, 2022.
doi: 10.1109/ACCESS.2021.3139323.

\bibitem{Zhao2021}
R. Zhao and S. Wang,
``A Review of Quantum Neural Networks: Methods, Models, Dilemma,''
\textit{arXiv preprint arXiv:2109.01840},
2021.

\bibitem{Schuld2014}
M. Schuld, I. Sinayskiy, and F. Petruccione,
``The Quest for a Quantum Neural Network,''
\textit{Quantum Information Processing},
vol. 13, pp. 2567--2586, 2014.

\bibitem{Beer2020}
K. Beer, D. Bondarenko, T. Farrelly,
T. J. Osborne, R. Salzmann,
D. Scheiermann, and R. Wolf,
``Training Deep Quantum Neural Networks,''
\textit{Nature Communications},
vol. 11, no. 1, p. 808, 2020.
doi: 10.1038/s41467-020-14454-2.



\end{thebibliography}
\end{document}